\documentclass[letterpaper, 10pt, conference]{ieeeconf}  

\IEEEoverridecommandlockouts    

\usepackage{graphicx}
\usepackage{amssymb,amsmath,bm,color}
\usepackage{textcomp}
\usepackage{balance}
\usepackage[table,xcdraw]{xcolor}
\def\vec#1{\ensuremath{\bm{{#1}}}}

\usepackage[justification=centering]{caption}

\newcommand{\etal}{\textit{et al.~}}

\usepackage{multirow}
\usepackage{algorithm}
\usepackage[noend]{algpseudocode}
\usepackage{color}

\makeatletter
\def\BState{\State\hskip-\ALG@thistlm}
\makeatother

\title{\LARGE \bf
Bag-of-Audio-Words based on Autoencoder Codebook for Continuous Emotion Prediction
}

\author{Mohammed Senoussaoui, Patrick Cardinal, Alessandro Lameiras Koerich
\thanks{*The authors acknowledge partial funding from FRQNT}
\thanks{Mohammed Senoussaoui, Patrick Cardinal and Alessandro Lameiras Koerich are with the Department of Software and IT Engineeing, \'{E}cole de Technologie Sup\'{e}rieure (\'{E}TS), Universit\'{e} du Qu\'{e}bec, Montreal, QC, Canada
    {\tt\small mohammed.senoussaoui.1@ens.etsmtl.ca, patrick.cardinal@etsmtl.ca, alessandro. lameiras-koerich@etsmtl.ca}}%
}

\begin{document}

\maketitle

\begin{abstract}
In this paper we present a novel approach for extracting a Bag-of-Words (BoW) representation based on a Neural Network codebook. The conventional BoW model is based on a dictionary (codebook) built from elementary representations which are selected randomly or by using a clustering algorithm on a training dataset. A metric is then used to assign unseen elementary representations to the closest dictionary entries in order to produce a histogram. In the proposed approach, an autoencoder (AE) encompasses the role of both the dictionary creation and the assignment metric. The dimension of the encoded layer of the AE corresponds to the size of the dictionary and the output of its neurons represents the assignment metric. Experimental results for the continuous emotion prediction task on the AVEC 2017 audio dataset have shown an improvement of the Concordance Correlation Coefficient (CCC) from 0.225 to 0.322 for arousal dimension and from 0.244 to 0.368 for valence dimension relative to the conventional BoW version implemented in a baseline system. 
\end{abstract}


\section{Introduction}
\label{sec:intro}
Human speech signal carries several levels of information. The most obvious one is the classical linguistic message that humans use to communicate with each other. Besides that, according to Scherer \etal \cite{Scherer2003}, at a different level, emotional information that can also be transmitted through a change in vocal parameters. In particular, the task of measuring the emotional state of a speaker continuously is very challenging and complex even if the literal meaning of his/her elocution is perfectly understood. Humans usually use distinct vocal cues to express their emotion. 

Affective computing is a relatively recent research field and its main aim is to change the classical interaction way between human beings and machines by attempting to embed into machines some of the natural human capacities \cite{picard1995affective}. In the last years, several works have attempted to predict continuous emotion states such as pleasure/displeasure and alertness/activeness based on the automatic analysis of multimodal information such as video, audio, ECG and other physiological measures \cite{Cardinal2015,trigeorgis2016adieu,chen2017multimodal,ringeval2017avec,schmitt2016border,cardinal2015ets,silva2019emotion}. In particular, a periodic challenge called Audio-Visual Emotion Challenge and Workshop (AVEC) has been organized since 2011 and its main goal has been the comparison of multimedia processing and machine learning methods for automatic audio, visual, and audiovisual emotion analysis and other related tasks. In AVEC 2017, one of the two proposed sub-challenges consisted of the continuous prediction of three emotion dimensions, namely arousal, valence and likability, from human to human real-life interaction, recorded "in-the-wild" \cite{ringeval2017avec}. The participants were challenged to use arbitrary quality audiovisual materials from SEWA dataset\footnote{https://sewaproject.eu}, recorded using personal equipment in order to predict the level of affect of the three emotion dimensions. The AVEC 2017 affect sub-challenge using only the audio modality represents the main research topic of this paper.

The winning team of the affect sub-challenge of AVEC 2017 \cite{chen2017multimodal} proposed an approach based on: (i) pre-trained neural network for 1D audio representation (Soundnet \cite{aytar2016soundnet}) as well as handcrafted acoustic features \cite{schuller2011recognising} such as Low Level Descriptors (LLD), and (ii) a Long Short Term Memory (LSTM) recurrent neural network as a regression model. On the other hand, from the purely theoretical standpoint, adopting powerful feature representations able to model most faithfully the signal is helpful for the subsequent prediction process. In the AVEC 2017 baseline system \cite{ringeval2017avec}, besides the LLD, which is used to capture local signal information in time, two other segment-level representations are also adopted, namely, functionals and BoW \cite{ringeval2017avec}. BoW is a common feature extraction procedure for text information in the field of Natural Language Processing (NLP) that looks at the histogram of the words within the text. This procedure was taken successfully by the audio and video processing communities \cite{schmitt2016border,fei2005bayesian,pancoast2013n}.

The main contribution of this paper is a novel approach to generate Bag-of-Audio-Words (BoAW) representation based on an autoencoder (AE). An AE is a type of artificial neural network that learns efficient data coding from unsupervised data. The AE encompasses simultaneously the two main steps of the BoW procedure, the creation of a vocabulary of known words and the measure of the presence of known words. The dimension of the AE encoded layer represents the size of the dictionary and the values of its neurons is related to the measure of the presence of the words. Experiments on continuous emotion prediction task using AVEC 2017 dataset have shown an improvement of the Concordance Correlation Coefficient (CCC) from 0.225 to 0.322 for arousal dimension and from 0.244 to 0.368 for valence dimension compared to the conventional BoW version implemented in the baseline system of AVEC 2017. Furthermore, the results achieved by the proposed BoAW approach for the arousal dimension (CCC = 0.497), without any score post-processing, are close to the best-known result (CCC = 0.526) for this task \cite{ringeval2017avec}, which nevertheless employs post-processing techniques to improve the final CCC.

This paper is organized as follows. Section \ref{sec:Bag-of-Words} reviews the original version of Bag-of-Words (BoW) signal representation. Section \ref{sec:Bag-of-Words-NN} describes the proposed version of BoW based on a neural network dictionary and assignment metric. The experimental protocol, experimental results as well as discussions on the results are presented in Sections \ref{sec:Experimental-Setup} and \ref{Results_Discussions} respectively. Finally, the conclusions and perspective of future directions are presented in the last section.  
\begin{figure*}[!htpb]
\centering
\includegraphics[width=0.8\textwidth,height=7cm]{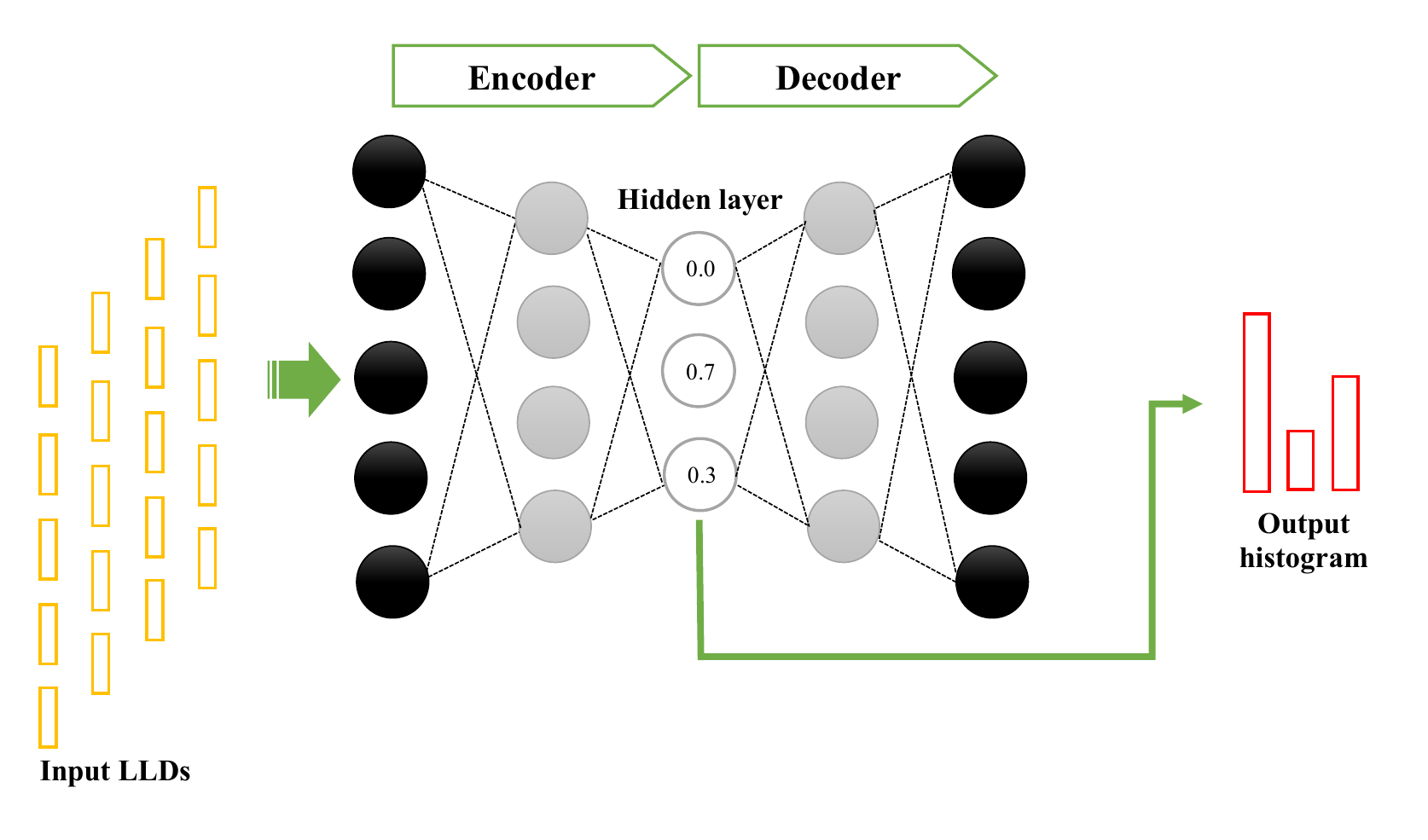}
\caption{Illustration of the architecture of the proposed Bag-of-Audio-Words based on a neural network codebook (BoAW-NN).}
\label{fig:BoW-NN}
\end{figure*}

\section{Bag-of-Words (BoW)}
\label{sec:Bag-of-Words}
The Bag-of-Words is a common representation of information in the Natural Language Processing (NLP) field. The main idea behind BoW is the use of a rich dictionary (i.e. codebook) of elementary words to represent a whole document based on word appearance frequencies to build a histogram. The main advantage of the BoW representation is its capacity of summarizing the meaningful information of a variable length input document using a fixed length vector (i.e. the histogram). Note that, the fixed length representation is more suitable for most of the machine learning algorithms. Nevertheless, the BoW representation depends on the management of the size of the vocabulary to avoid sparse representations. 

Lately, the BoW representation has been extended to model other data modalities such as image and video under the name of Bag-of-Visual-Words (BoVW) and more recently to model audio data under the name Bag-of-Audio-Words (BoAW) \cite{schmitt2016border,fei2005bayesian,pancoast2013n,schmitt2017openxbow}. Using modalities other than text requires the definition of elementary units "equivalent" to lexical words, e.g. pixels, contours in the case of an image or short-term frames in the case of an audio signal.

The process of building a BoW extractor consists of two main components:
\begin{itemize}

\item \textbf{Building a Vocabulary (or Dictionary):} it represents the vocabulary (or codebook) of the most important elementary representations, which play the role of anchor models in the studied phenomena \cite{sturim2001speaker}. This codebook can be generated in several ways, starting from a simple random assignation of codewords and ending into a more sophisticated clustering model such as K-means or Gaussian Mixture Model (GMM) \cite{schmitt2017openxbow}.

\item \textbf{Assignment Mechanism:} for a given document (image, video, audio, etc., each word is assigned to a dictionary entry (codeword) and the appearance frequencies of the whole entries constitute the output histogram. Unlike the conventional text BoW, in which the comparison between words is obvious, the assignment of the elementary representations in image (BoVW) or audio (BoAW) models, needs the adoption of a specific metric. The choice of this metric is quite related to the choice of the dictionary generation method. This is in the sense that the Euclidean distance is more suitable for random-selection method and for K-means, while stochastic distances are used for the GMM model.   
\end{itemize} 

The building of a dictionary and the definition of an assignment mechanism constitute the core of a BoW extractor. Therefore, the improvement of these two components may certainly lead to improvements in the whole model. In the next section we propose a strategy based on autoencoders that learn a dictionary and the assignment metric simultaneously.

\section{Bag-of-Words based on Neural Network Codebook (BoW-NN)}
\label{sec:Bag-of-Words-NN}
The main focus in this research is on the improvement of the conventional version of Bag-of-Audio-Words (BoAW), which use random sampling or clustering algorithms for codebook generation and term frequencies for generating the bag-of-audio-words. To such an aim we propose the adoption of an autoencoder (AE) to build a codebook. An autoencoder is a special type of multi-layer neural network trained to reproduce its input at its output and internally it has a hidden layer (encoded) that describes the code used to represent the input.

The main idea consists of using the neurons of the hidden layer as anchors (i.e. the codewords) and their outputs as the assignment probabilities of a given acoustic frame as illustrated in Fig.~\ref{fig:BoW-NN}. This is motivated by the evidence that for similar inputs, a well-trained AE should be able to activate similar neuronal regions of its hidden layer. Since an AE creates a reduced representation of the data, it is a natural approach for creating dictionaries. This mechanism is quite similar to the clustering one used to generate the BoW dictionaries. 

Using an AE to generate a BoAW has the advantage of encompassing the roles of both BoW components i.e. building a dictionary together with an embedded assignment mechanism. Furthermore, the neural network model gives the flexibility of using different low-level descriptors as input, including raw signal, spectrogram image, sequential data, etc. Finally, the activation probabilities computed by the encoded layer facilitate the task of implementing different assignment strategies, such as single assignment, multiple assignments or soft assignment, which are described as follows:

\begin{itemize}
\item \textbf{Single Assignment Strategy} in this strategy each acoustic frame is assigned to the closest codeword in the codebook. Indeed, we do not use this strategy because it is too restrictive; 

\item \textbf{Multiple Assignments Strategy} in this strategy, each acoustic frame is assigned to the closest $N$ words in the codebook in order to calculate the frequencies of the output histogram. Therefore, in this strategy we need to choose $N$ appropriately;

\item \textbf{Soft Assignment Strategy} unlike the previous strategy, in which a fixed number of words is used ($N$ words), in this strategy we propose to use a variable number of closest words governed by a scalar threshold $\theta$. Thus, a given acoustic frame will be assigned to all words in which the assignment probability is greater or equal to $\theta$. Note that a normalization of the Euclidean length (norm) of the vector representing the histogram was found to be effective to improve the final regression performance.    
\end{itemize}

\section{Proposed Approach}
\label{sec:Approach}

\begin{figure*}[!htpb]
\centering
\includegraphics[width=0.7\textwidth]{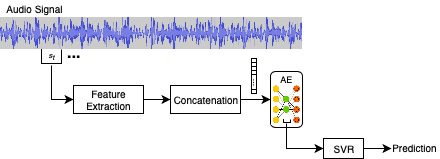}
\caption{An overview of the proposed approach.}
\label{fig:over}
\end{figure*}

The proposed approach starts with a sliding window that captures part of the audio signal and extracts low level features from it. However, before going to the AE, a feature vector is concatenated with the seven previous and posterior ones to make up a 365-dimensional vector. The AE then generates a codeword at its encoded layer. Such a codeword is used to train a support vector regressor (SVR), during the training step or to predict a certain level of valence, arousal or liking during the test step. The proposed approach is illustrated in Fig.~\ref{fig:over}. 

\subsection{Feature Extraction and Concatenation}
   
We use the extended Geneva Minimalistic Acoustic Parameter Set (eGeMAPS) \cite{eyben2016geneva}, which are low-level descriptors (LLD) that cover spectral, cepstral, prosodic and voice quality information of the voice record. Such features have been used in the RECOLA baseline together with other modalities \cite{Valstar2016}. A 23-dimensional acoustic low-level descriptor is extracted every 10 ms over a short-term frame. Furthermore, we concatenate 15 LLD frames, the current one with its seven past frames and its seven forthcoming frames as well.

In the case of the baseline, high-level acoustic descriptors (HLD) are computed from the LLD over segments of six seconds with a shift of 100 ms. For the baseline BoAW feature, a codebook of 1,000 codewords was used to create the histogram of audio words. For the proposed BoAW-NN approach, on the other hand, the number of words corresponds to the number of neurons in the AE encoded (hidden) layer. This number is empirically optimized on the development set to be equal to 345.

\subsection{Autoencoder Configuration}
An autoencoder with one input layer, two hidden layers and an output layer is trained on the training subset using Tensorflow with Keras interface. The rectified linear unit function ("ReLU") with a maximum value equal one was used as activation and "RMSprop" is used as optimization algorithm. The number of input neurons is fixed to 345, which is equal to the 15 LLD frames, each of 23 dimensions. The number of neurons in the two hidden layers has been fixed empirically to be equal to the number of the inputs of the input layer i.e. 345. 

\subsection{Regression Model}
Support vector regressors (SVR) with linear kernel have been shown to give competitive accuracy for problems where the number of features is very large as well as for problems where the number of features is much larger than the number of instances. In such cases, a nonlinear mapping usually does not improve the performance. Linear SVRs are good enough and they enjoy much faster training/testing times.

Given a set of training vectors $\textbf{x}$ and the respective target values $\textbf{y}$, where $\textbf{x}\in\Re^n$ and $\textbf{y}\in\Re$, the linear SVR finds a model $\textbf{w}$ such that $\textbf{w}^\top\textbf{x}$ is close to the target value $\textbf{y}$. It solves the following regularized optimization problem.

\[ \min_{\textbf{w}} f(\textbf{w})\equiv\frac{1}{2}\textbf{w}^\top\textbf{w}+C\sum\xi_\epsilon (\textbf{w};\textbf{x},\textbf{y}) \]
where $C>0$ is the regularization parameter and

\[ \xi_\epsilon (\textbf{w};\textbf{x},\textbf{y}) = \max(|\textbf{w}^\top\textbf{x}-\textbf{y}|-\epsilon,0) \]
\noindent is the $\epsilon$-insensitive loss function associated with $(\textbf{x},\textbf{y})$.

In this work we used the linear SVR implementation provided with the LIBLINEAR tool \cite{REF08a}.

\section{Experimental Results}
\label{sec:Experimental-Setup}
For the evaluation of the proposed BoAW-NN approach, we have carried out several experiments on the dataset of the affect sub-challenge of AVEC 2017 challenge. The dataset used in this challenge is a subset of German subjects taken from the Sentiment Analysis in the Wild (SEWA) dataset. SEWA dataset consists of audiovisual recordings of spontaneous and natural interactions. Each recording contains a target subject and his/her interlocutor, and it was collected using personal computer webcams and microphones from offices or from homes.

In total 32 pairs, i.e. 64 subjects are provided to the participants of the challenge. The whole subset was broken-down into three partitions (training, development, and test), where both a speaker and an interlocutor appear into a single partition, meaning that there is no overlapping between speakers and interlocutors among data partitions. Speaker turn timings were also provided in order to know if either a speaker or an interlocutor is speaking at a given moment. This is a useful information since our target is to predict the emotion of speakers and not of interlocutors. 

\subsection{Evaluation Protocol}
The participants in the affective sub-challenge were required to perform fully continuous affect recognition (each 100 ms, the affect level has to be predicted) of three affective dimensions: arousal, valence, and likability. The official competition performance metric is the Concordance Correlation Coefficient (CCC), denoted in Equation~\ref{CCC}. It combines the Pearson correlation coefficient with the squared difference between the means of the prediction ($y$) and the labels ($\hat{y}$). This metric measures the association between variables and penalizes the score even if the model predicts the emotion well but shifts the value.

\begin{equation}
\rho_{c}=\frac{2\rho\sigma_{y}\sigma_{\hat{y}}}{\sigma_{y}^{2}+\sigma_{\hat{y}}^{2}+(\mu_{y}-\mu_{\hat{y}})^{2}}
\label{CCC}
\end{equation}

\noindent where $\rho$ is the Pearson correlation coefficient between the label and the gold standard $\sigma^{2}$ is the variance, $\mu_{y}$ and $\mu_{\hat{y}}$ are the mean values of the prediction and the gold standard.

A main constraint of the AVEC 2017 challenge is that the participants do not have access to the golden standard of the test set. Nevertheless, they have up to five evaluation trails in the test set during the challenge and they have up to 10 trials after the end of the challenge. The performance evaluation on the test set is performed only by the challenge organizers, based on the predictions submitted by the participants. 

\subsection{Speaker Turn Information}
For each recording of the SEWA dataset used in AVEC 2017 challenge, the information about speaker turn is provided. It is important to mention that we are interested in predicting the emotion of the speaker, while he/she answers several questions posed by an interlocutor. Chen \etal \cite{chen2017multimodal} have proposed three strategies to include the speaker turn information in the regression model: mixed, purified and doubled. Besides these three strategies, we have also proposed a novel strategy called "as feature".

\begin{itemize}
\item \textbf{Mixed}: This is the basic case, in which there is no distinction between the speech of the speaker of interest and the speech of its interlocutor. The speech of both is used without any information of speaker turns.
\item \textbf{Purified}: In this strategy only the acoustic speech frames of the target speaker are used. The other frames are replaced by zeros. 
\item \textbf{Doubled}: As its name indicates, the size of the acoustic frames is doubled by padding a vector of zeros to each frame. For the target speaker frames, the zeros are padded after the frames while for the interlocutor the zeros are before the frames.  
\item \textbf{As Feature}: This strategy simply adds an additional bit of information to the acoustic frames to indicate the speaker turn.
\end{itemize}


\begin{table*}[ht]
\centering
\caption{Results on the development set as measured by the Concordance Correlation Coefficient (CCC). The results are broken down by four strategies for including speaker turn and three score scaling methods.}
\label{tab:ResDev}
\def\arraystretch{1.5}
\begin{tabular}{|l|l|c|c|c|c|c|c|c|c|c|}
\hline
\multicolumn{1}{|c|}{} & \multicolumn{1}{c|}{} & \multicolumn{3}{c|}{\textbf{Arousal}} & \multicolumn{3}{c|}{\textbf{Valence}} & \multicolumn{3}{c|}{\textbf{Liking}} \\ \cline{3-11} 
\multicolumn{1}{|c|}{\multirow{-2}{*}{\textbf{Features}}} & \multicolumn{1}{c|}{\multirow{-2.5}{*}{\textbf{\begin{tabular}[c]{@{}l@{}}Speaker\\Turn\\Info\end{tabular}}}} & \multicolumn{1}{c|}{\textit{\begin{tabular}[c]{@{}c@{}}No\\Scaling\end{tabular}}} & \multicolumn{1}{c|}{\textit{\begin{tabular}[c]{@{}c@{}}SD Ratio \\ Scaling\end{tabular}}} & \multicolumn{1}{c|}{\textit{\begin{tabular}[c]{@{}c@{}}Min-Max \\ Scaling\end{tabular}}} & \multicolumn{1}{c|}{\textit{\begin{tabular}[c]{@{}c@{}}No\\Scaling\end{tabular}}} & \multicolumn{1}{c|}{\textit{\begin{tabular}[c]{@{}c@{}}SD Ratio \\ Scaling\end{tabular}}} & \multicolumn{1}{c|}{\textit{\begin{tabular}[c]{@{}c@{}}Min-Max \\ Scaling\end{tabular}}} & \multicolumn{1}{c|}{\textit{\begin{tabular}[c]{@{}c@{}}No\\Scaling\end{tabular}}} & \multicolumn{1}{c|}{\textit{\begin{tabular}[c]{@{}c@{}}SD Ratio \\ Scaling\end{tabular}}} & \multicolumn{1}{c|}{\textit{\begin{tabular}[c]{@{}c@{}}Min-Max \\ Scaling\end{tabular}}} \\ \hline
\begin{tabular}[c]{@{}l@{}}Baseline AVEC 2017\\(BoAW) \end{tabular} & \textbf{Mixed} & 0.343 & 0.376 & 0.456 & 0.390 & 0.420 & 0.357 & 0.074 & 0.078  & 0.054 \\ \hline
 & \textbf{Mixed} & 0.448 & 0.454 & 0.273 & 0.419 & 0.437 & 0.337 & 0.098 & 0.085  & 0.085 \\ \cline{2-11} 
 & \textbf{Purified}  & 0.342 & 0.398 & 0.368 & 0.283 & 0.347 & 0.328 & \textbf{0.144} & \textbf{0.150}  & \textbf{0.147} \\ \cline{2-11} 
 & \textbf{Doubled} & 0.469 & 0.470 & 0.329 & 0.405 & 0.424 & 0.259 & 0.104 & 0.095  & 0.141 \\ \cline{2-11} 
\multirow{-4}{*}{\begin{tabular}[c]{@{}l@{}}BoAW-NN \\ (Multiple Assignments:\\ $N=20$).\end{tabular}}  & \textbf{As Feature}  & 0.461 & 0.467 & 0.278 & 0.414 & 0.435 & 0.303 & 0.112 & 0.100  & 0.085 \\ \hline
 & \textbf{Mixed}  & 0.463 & 0.459 & 0.550 & \textbf{0.442} & \textbf{0.468} & \textbf{0.433} & 0.093 & 0.083  & 0.132 \\ \cline{2-11} 
 & \textbf{Purified}  & 0.382 & 0.414 & 0.497 & 0.288 & 0.368 & 0.321 & 0.126 & 0.137  & 0.142 \\ \cline{2-11} 
 & \textbf{Doubled} & \textbf{0.497}  & \textbf{0.502} & \textbf{0.577} & 0.430 & 0.453 & 0.416 & 0.115 & 0.113  & 0.15  \\ \cline{2-11} 
\multirow{-4}{*}{\begin{tabular}[c]{@{}l@{}}BoAW-NN\\ (Soft Assignment:\\ $\theta=0.05$)\end{tabular}} & \textbf{As Feature}  & 0.264 & 0.239 & 0.307 & 0.269 & 0.281 & 0.314 & 0.062 & 0.042  & 0.106 \\ \hline
\end{tabular}
\end{table*}

\subsection{Regression Score Scaling}

The prediction process implements a series of information transformations, starting from the input raw signal to finish with the predicted scores. Although, all these operations have the same goal of filtering the undesirable variability in the signal or in its intermediate representations, some of this variability can still be found into the predicted scores. Scaling these scores could help to attenuate some of the remaining noise. Thus, we have adopted two methods to scale the output scores:  
\begin{itemize}
\item \textbf{Standard-Deviation Ratio}:  This scaling method has already been used by Trigeorgis \etal \cite{trigeorgis2016adieu} and it has been shown interesting results. A normalized output vector $\vec{y}_{norm}$ is calculated using Equation \ref{eq:std_ratio}:

\begin{equation} 
\label{eq:std_ratio}
\vec{y}_{norm} = \frac{\sigma_{p}}{\sigma_{l}}\otimes \vec{y}
\end{equation}
\noindent where $\sigma_{p}$ is the standard deviation of the predictions, $\sigma{l}$ is the standard deviation of the golden standard, $\otimes$ is the element-wise multiplication operation and $\vec{y}$ is the prediction vector to be scaled. 

\item \textbf{Min-Max Scaling}: This scaling provides a linear transformation of the original range of predictions in a pre-defined boundary. A normalized output vector $\vec{y}_{norm}$ is calculated using Equation \ref{eq:minMax_ratio}:

\begin{equation} 
\label{eq:minMax_ratio}
\vec{y}_{norm} = \frac{\max_{l}-\min{l} \cdot \vec{y}-\min_{p}}{\max_{p}-\min_{p}}+\min_{l} 
\end{equation}

\noindent where $\max_{l}$ and $\min_{l}$ are the maximum and minimum values in the training labels respectively, and $\max_{p}$ and $\min_{p}$ are the maximum and minimum values of $\vec{y}$, the original prediction vector, respectively.  
\end{itemize}

\subsection{Results and Discussion}
\label{Results_Discussions}
Table \ref{tab:ResDev} shows the results achieved on the development set of AVEC 2017 affect sub-challenge using the support vector regressor (SVR) with a linear kernel \cite{REF08a}. First, in the case where no score scaling method is used, we can see that both of the proposed versions of BoAW-NN (i.e. with \textit{multiple} and with \textit{soft assignment}) outperform the performance of the original BoAW used in the AVEC 2017 baseline system. For the arousal dimension, the BoAW-NN with \textit{soft-assignment} using a threshold parameter $\theta$ = $0.05$ and the \textit{Doubled} speaker turn strategy shows the best performance, achieving a CCC of 0.497, compared to a CCC of 0.469 achieved by the BoAW-NN with \textit{multiple assignments} and to a CCC of 0.343 achieved by the original BoAW baseline version. For the valence dimension, the best CCC of 0.442 is again achieved by BoAW-NN with \textit{soft-assignment}, while the BoAW-NN with \textit{multiple assignments} achieved a CCC of 0.419 and the baseline version achieved a CCC of only 0.390. Finally, for the liking dimension, it was the of BoAW-NN with \textit{multiple assignments} that achieved the best result with a CCC of 0.144 compared to a CCC of 0.126 achieved by the BoAW-NN with \textit{soft-assignment} and a CCC of 0.074 for the baseline version. However, these are not the best overall results at all. Table \ref{tab:ResDev} shows that the score scaling methods are generally helpful and improve the CCC. The \textit{Min-Max Scaling} was very effective for the arousal dimension, in which a CCC of 0.577 is achieved, compared to a CCC of 0.497 without normalization. The \textit{Standard-Deviation Ratio} works better for the valence and liking dimensions and also allows an improvement in the CCC values.  
Table \ref{tab:ResTest} shows the results achieved on the test set\footnote{Given that the results for the baseline BoAW are available without any score scaling, and the limited number of trials on the test set, we present only the results without score scaling.} of AVEC 2017 affect sub-challenge and their corresponding results on the development set. These results are obtained without any score scaling. Again, the BoAW-NN approach outperforms the original BoAW on the three affective dimensions. Note that the proposed BoAW-NN approach also has the advantage of being more compact in terms of dimension (1,000 for BoAW vs. 345 for BoAW-NN) over the original BoAW. 

\begin{table}[ht!]
\caption{BoAW-NN results on the test set (Test-), as measured by the concordance correlation coefficient (CCC). The corresponding results on the development set (Dev-) are also provided.}
\label{tab:ResTest}
\centering
\def\arraystretch{1.5}%
\begin{tabular}{|l|c|c|c|}
\hline
\textbf{Approach} & \multicolumn{1}{l|}{\textbf{Arousal}} & \multicolumn{1}{l|}{\textbf{Valence}} & \multicolumn{1}{l|}{\textbf{Liking}} \\ \hline 
Dev-Baseline  & 0.344 & 0.351 & 0.081 \\ \hline 
Dev-NN-BoAW & \bf{0.497} & \bf{0.442} & \bf{0.150}  \\ \hline
Test-Baseline & 0.225 & 0.244 & -0.020 \\ \hline
Test-NN-BoAW  & \bf{0.322} & \bf{0.368} & \bf{0.096}  \\ \hline
\end{tabular}
\end{table}

\section{Conclusion}
In this paper, we present a novel approach for building a Bag-of-Audio-Words (BoAW) based on an autoencoder (AE) dictionary. From unsupervised data, the AE learns to encode the acoustics frames using its encoded layer. The neurons of that layer simulate the role of the dictionary words (codewords) while their activation values measure the assignment probability of a given frame to a given word. Based on the assignment strategy, two variants of BoAW-NN have been presented, namely, \textit{multiple} and with \textit{soft assignment}. 

The proposed approach was evaluated on the audio data of the AVEC 2017 affect sub-challenge and compared with the baseline BoAW. The proposed BoAW-NN with its two variants, outperforms the original version on both the development and the test sets. The CCCs achieved by the proposed approach are 45\%, 14\% and 95\% higher than CCCs provided by the baseline BoAW approach for arousal, valence and liking respectively. The promising results encourage us to continue exploring the BoAW-NN approach. In the future, we are planning to replace the autoencoder by a supervised Deep Neural Network (DNN) with a bottleneck layer \cite{Yu2014,richardson2015}. 


\balance

\bibliographystyle{IEEEtran}

\end{document}